\documentclass{article}
\usepackage{amssymb}
\usepackage{amsmath}
\usepackage{amsfonts}

\newtheorem{theorem}{Theorem}

\newtheorem{lemma}[theorem]{Lemma}

\input{tcilatex}

\begin{document}

\begin{center}
{\LARGE On the Liouville-Arnold integrable flows related with quantum
algebras and their Poissonian representations}\bigskip

\textbf{A. M. Samoilenko}$^{\ast )}$\textbf{, Y.A. Prykarpatsky}$^{\ast
)\ast \ast )}$\textbf{, D.L. Blackmore}$^{\ast \ast \ast )}$ \textbf{and}  
\textbf{A. K. Prykarpatsky}$^{\ast \ast )}$,

*) Institute of Mathematics at the National Academy of Sciences, 3
Tereshchenkivska Str., Kyiv 00601 Ukraina

**) Dept. of Applied Mathematics at the AGH University of Science and
Technology, 30 Mickiewicz Al. Bl. A4, 30059 Krakow, Poland and Dept. of
Nonlinear Math. Analysis at the Institute of APMM of the Nat. Acad. of
Sciences,\ Lviv,79601(email: prykanat@cyberagl.com, pryk.anat@ua.fm)

***) Dept. of Mathem. Studies at the NJIT, University Heights, New Jersey
07102 USA
\end{center}

\ \ \ \ \ \ \ \ \ \ \ \ \ \ \ \ \ \ \ \ \ \ \ \ \ \ \ \ \ \ \ \ \ \ \ \ \ \
\ \ \ \ \ \ \ \ \ \ \ \ \ \ \ \ \ \ \ \ \ \ \ \ \ \ \ \ \ \ \ \ \ \ \ \ \ \
\ \ \ \ \ \ \ \ \ \ \ \ \ \ \ \ \ \ \ \ \ \ \ \ \ \ 

\textbf{Abstract.} Based on the structure of Casimir elements associated
with general Hopf algebras there are constructed Liouville-Arnold integrable
flows related with naturally induced Poisson structures on \ arbitrary \
co-algebra\ \ \ \ and their deformations.\ \ \ Some interesting special
cases including the oscillatory Heisenberg-Weil algebra related co-algebra
structures and adjoint with them\ \ \ \ integrable Hamiltonian systems are
considered.\ \ 

\section{Hopf algebras and co-algebras: main definitions}

\setcounter{equation}{0} \renewcommand{\theequation}{\arabic{section}.%
\arabic{equation}}Consider a Hopf algebra $\mathcal{A}$ over $\mathbb{C}$
endowed with two special homomorphisms called coproduct $\Delta :\mathcal{%
A\rightarrow A\otimes A}$ and counit $\varepsilon :\mathcal{A\rightarrow }%
\mathbb{C},$ as well an antihomomorphism (antipode) $\nu :\mathcal{%
A\rightarrow A}$, such that for any $a\in \mathcal{A}$ 
\begin{eqnarray}
(id\otimes \Delta )\Delta (a) &=&(\Delta \otimes id)\Delta (a),  \label{1.1}
\\
(id\otimes \varepsilon )\Delta (a) &=&(\varepsilon \otimes id)\Delta (a)=a, 
\notag \\
m((id\otimes \nu )\Delta (a)) &=&m((\nu \otimes id)\Delta (a))=\varepsilon
(a)I,  \notag
\end{eqnarray}%
where $m:\mathcal{A\otimes A\rightarrow A}$ is the usual multiplication
mapping, that is for any $a,b\in \mathcal{A}$ $m(a\otimes b)=ab.$ The
conditions (\ref{1.1}) were introduced by Hopf $\cite{Ho}$ in a
cohomological context. Since most of the Hopf algebras properties depend on
the coproduct operation $\Delta :\mathcal{A\rightarrow A\otimes A}$ and
related with it Casimir elements, below we shall dwell mainly on the objects
called co-algebras endowed with this coproduct.

The most interesting examples of co-algebras are provided by the universal
enveloping algebras $U(\mathcal{G})$ of Lie algebras $\mathcal{G}.$ If, for
instance, a Lie algebra $\mathcal{G}$ possesses generators $X_{i}\in 
\mathcal{G},$ $i=\overline{1,n},$ $n=\dim \mathcal{G},$ the corresponding
enveloping algebra $U(\mathcal{G)}$ can be naturally endowed with a Hopf
algebra structure by defining 
\begin{eqnarray}
\Delta (X_{i}) &=&I\otimes X_{i}+X_{i}\otimes I,\text{ }\Delta (I)=I\otimes
I,  \label{1.2} \\
\varepsilon (X_{i}) &=&-X_{i},\text{ \ \ \ \ \ \ \ \ \ \ \ }\nu (I)=-I. 
\notag
\end{eqnarray}%
These mappings acting only on the generators of $\mathcal{G}$ are
straightforwardly extended to any monomial in $U(\mathcal{G)}$ by means of
the homomorphism condition $\Delta (XY)=\Delta (X)\Delta (Y)$ for any $%
X,Y\in \mathcal{G\subset }U(\mathcal{G)}.$ In general\TEXTsymbol{<} an
element $Y\in U(\mathcal{G)}$ of a Hopf algebra such that $\Delta
(Y)=I\otimes Y+Y\otimes I$ is called primitive, and the known Friedrichs
theorem $\cite{Po}$ ensures, that in $U(\mathcal{G)}$ the only primitive
elements are exactly generators $X_{i}\in \mathcal{G},$ $\ i=\overline{1,n}.$

On the other handside, the homomorphism condition for the coproduct $\Delta :%
\mathcal{A\rightarrow A\otimes A}$ implies the compatibility of the
coproduct with the Lie algebra commutator structure:%
\begin{equation}
\lbrack \Delta (X_{i}),\Delta (X_{j})]_{\mathcal{A\otimes A}}=\Delta
([X_{i},X_{j}]_{\mathcal{A}})  \label{1.3}
\end{equation}%
for any $X_{i},X_{j}\in \mathcal{G},$ \ $i,j=\overline{1,n}.$ Since the
Drinfeld report $\cite{Dr}$ the co-algebras defined above are also often
called "quantum" groups due to their importance $\cite{KBI}$ in studying
many two-dimensional quantum models of modern field theory and statistical
physics.

It was also observed (see for instance $\cite{KBI}),$ that the standard
co-algebra structure (\ref{1.2}) of the universal enveloping algebra $U(%
\mathcal{G)}$ can be nontrivially extended making use of some its
infinitesimal deformations\TEXTsymbol{<} saving the co-associativity (\ref%
{1.3}) of the deformed coproduct $\Delta :U_{z}(\mathcal{G)\rightarrow }U_{z}%
\mathcal{(\mathcal{G)}\otimes }U_{z}\mathcal{(\mathcal{G)}}$ with $U_{z}%
\mathcal{(\mathcal{G)}}$ being the corresponding universal enveloping
algebra deformation by means of a parameter $z\in \mathbb{C},$ such that $%
\underset{z\rightarrow 0}{lim}U_{z}\mathcal{(\mathcal{G)=}}U\mathcal{(%
\mathcal{G)}}$ subject to some natural topology on $U_{z}\mathcal{(\mathcal{%
G)}}.$

\section{Casimir elements and their special properties}

\setcounter{equation}{0}Take any Casimir element $C\in U_{z}\mathcal{(%
\mathcal{G)}},$ that is an element satisfying the condition $[C,U_{z}%
\mathcal{(\mathcal{G)]=}}0,$ and consider the action on it of the coproduct
mapping $\Delta :$%
\begin{equation}
\Delta (C)=C(\{\Delta (X)\}),  \label{2.1}
\end{equation}%
where we put, by definition, $C:=C(\{X\})$ with a set $\{X\}\subset \mathcal{%
\mathcal{G}}.$ It is a trivial consequence that for $\mathcal{A}:=U_{z}%
\mathcal{(\mathcal{G)}}$%
\begin{equation}
\lbrack \Delta (C),\Delta (X_{i})]_{\mathcal{A\otimes A}}=\Delta ([C,X_{i}]_{%
\mathcal{A}})=0  \label{2.2}
\end{equation}%
for any $X_{i}\in \mathcal{G},$ $\ i=\overline{1,n}.$

Define now recurrently the following $N-th$ coproduct $\Delta ^{(N)}:%
\mathcal{A\rightarrow }\overset{(N+1)}{\mathcal{\otimes }}\mathcal{A}$ for
any $N\in \mathbb{Z}_{+},$ where $\Delta ^{(2)}:=\Delta $ and $\Delta
^{(1)}:=id$ and 
\begin{equation}
\Delta ^{(N)}:=((id\otimes )^{N-2}\otimes \Delta )\cdot \Delta ^{(N-1)},
\label{2.3}
\end{equation}%
or as 
\begin{equation}
\Delta ^{(N)}:=(\Delta \otimes (id\otimes )^{N-2}\otimes id\otimes id)\cdot
\Delta ^{(N-1)}.  \label{2.4}
\end{equation}%
One can straightforwardly verify that 
\begin{equation}
\Delta ^{(N)}:=(\Delta ^{(m)}\otimes \Delta ^{(N-m)})\cdot \Delta
\label{2.5}
\end{equation}%
for any $m=\overline{0,N},$ and the mapping $\Delta ^{(N)}:\mathcal{%
A\rightarrow }\overset{(N+1)}{\mathcal{\otimes }}\mathcal{A}$ is an algebras
homomorphism, that is 
\begin{equation}
\lbrack \Delta ^{(N)}(X),\Delta ^{(N)}(Y)]_{\overset{(N+1)}{\mathcal{\otimes 
}}\mathcal{A}}=\Delta ^{(N)}([X,Y]_{\mathcal{A}})  \label{2.6}
\end{equation}%
for any $X,Y\in \mathcal{A}.$ In a particular case if $\mathcal{A=}U\mathcal{%
(\mathcal{G),}}$ the following exact expression 
\begin{eqnarray}
\Delta ^{(N)}(X) &=&X(\otimes id)^{N-1}\otimes id+id\otimes X(\otimes
id)^{N-1}\otimes id+...  \label{2.7} \\
&&...+(\otimes id)^{N-1}\otimes id\otimes X  \notag
\end{eqnarray}%
holds for any $X\in \mathcal{\mathcal{G}}.$

\section{Poisson co-algebras and their realizations}

\QTP{Body Math}
\bigskip \setcounter{equation}{0}As is well known $\cite{Pe},\cite{PM},$ a
Poisson algebra $\mathcal{P}$ is a vector space endowed with a commutative
multiplication and a Lie bracket $\{.,.\}$ including a derivation on $%
\mathcal{P}$ in the form 
\begin{equation}
\{a,bc\}=b\{a,c\}+\{a,b\}c  \label{3.1}
\end{equation}%
for any $a,b$ and $c\in \mathcal{P}.$ If $\mathcal{P}$ and $\mathcal{Q}$ are
Poisson algebras one can naturally define the following Poisson structure on 
$\mathcal{P}$ $\otimes \mathcal{Q}:$%
\begin{equation}
\{a\otimes b,c\otimes d\}_{\mathcal{P}\otimes \mathcal{Q}}=\{a,c\}_{\mathcal{%
P}}\otimes (bd)+(ac)\otimes \{b,d\}_{\mathcal{Q}}  \label{3.2}
\end{equation}%
for any $a,c\in \mathcal{P}$ \ and $b,d\in \mathcal{Q}.$ We shall also say
that $(\mathcal{P}$ $;\Delta )$ is a Poisson co-algebra if $\mathcal{P}$ \
is a Poisson algebra and $\Delta :\mathcal{P}$ $\mathcal{\rightarrow P}$ $%
\otimes \mathcal{P}$ is a Poisson algebras homomorphism, that is 
\begin{equation}
\{\Delta (a),\Delta (b)\}_{\mathcal{P}\otimes \mathcal{P}}=\Delta (\{a,b\}_{%
\mathcal{P}})  \label{3.3}
\end{equation}%
for any $a,b\in \mathcal{P}.$

\QTP{Body Math}
It is useful to note here that any Lie algebra $\mathcal{G}$ generates
naturally a Poisson co-algebra $(\mathcal{P};\Delta )$ by defining a Poisson
bracket on $\mathcal{P}$ by means of the following expression: for any $%
a,b\in \mathcal{P}$%
\begin{equation}
\{a,b\}_{\mathcal{P}};=<grad,\vartheta gradb>.  \label{3.4}
\end{equation}%
Here $\mathcal{P}$ $\simeq C^{\infty }(\mathbb{R}^{n};\mathbb{R})$ is a
space of smooth mappings linked with a base variables of the Lie algebra $%
\mathcal{G},$ $n=\dim \mathcal{G},$ and the implectic $\cite{PM}$ matrix $%
\vartheta :T^{\ast }(\mathcal{P})\rightarrow T(\mathcal{P})$ is given as 
\begin{equation}
\vartheta (x)=\{\tsum\limits_{k=1}^{n}c_{ij}^{k}x_{k}:i,j=\overline{1,n}\},
\label{3.5}
\end{equation}%
where $c_{ij}^{k},$ $i,j,k=\overline{1,n},$ are the corresponding structure
constants of the Lie algebra $\mathcal{G}$ and $x\in \mathbb{R}^{n}$ are the
corresponding linked coordinates. It is easy to check that the coproduct
(1.2) is a Poisson algebras homomorphism between $\mathcal{P}$ and $\mathcal{%
P}$ $\otimes \mathcal{P}.$ If one can find a "quantum" deformation $U_{z}(%
\mathcal{G)},$ then the corresponding Poisson co-algebra $\mathcal{P}_{z}$
can be constructed making use of the naturally deformed implectic matrix $%
\vartheta _{z}:T^{\ast }(\mathcal{P}_{z})\rightarrow T(\mathcal{P}_{z}).$
For instance, if $\mathcal{G=}so(2,1),$ there exists a deformation $%
U_{z}(so(2,1)\mathcal{)}$ defined by the following deformed commutator
relations with a parameter $z\in $ $\mathbb{C}:$%
\begin{eqnarray}
\lbrack \tilde{X}_{2},\tilde{X}_{1}] &=&\tilde{X}_{3},[\tilde{X}_{2},\tilde{X%
}_{3}]=-\tilde{X}_{1},  \label{3.6} \\
\lbrack \tilde{X}_{3},\tilde{X}_{1}] &=&\frac{1}{z}\sinh (z\tilde{X}_{2}), 
\notag
\end{eqnarray}%
where at $z=0$ elements $\left. \tilde{X}_{i}\right\vert _{z=0}=X_{i}\in
so(2,1),$ $i=\overline{1,3},$ compile \ a base of generators of the Lie
algebra $so(2,1).$ Then, based on expressions (\ref{3.6}) one can easily
construct the corresponding Poisson co-algebra $\mathcal{P}_{z},$ endowed
with the implectic matrix 
\begin{equation}
\vartheta _{z}(\tilde{x})=\left( 
\begin{array}{ccc}
0 & -\tilde{x}_{3} & -\frac{1}{z}\sinh (z\tilde{x}_{2}) \\ 
\tilde{x}_{3} & 0 & -\tilde{x}_{1} \\ 
\frac{1}{z}\sinh (z\tilde{x}_{2}) & \tilde{x}_{1} & 0%
\end{array}%
\right)  \label{3.7}
\end{equation}%
for any point $\tilde{x}\in \mathbb{R}^{3},$ linked naturally with the
deformed generators $\tilde{X}_{i},$ $i=\overline{1,3},$ taken above. Since
the corresponding coproduct on $U_{z}(so(2,1)\mathcal{)}$ acts on this
deformed base of generators as 
\begin{eqnarray}
\Delta (\tilde{X}_{2}) &=&I\otimes \tilde{X}_{2}+\tilde{X}_{2}\otimes I,
\label{3.8} \\
\Delta (\tilde{X}_{1}) &=&\exp (-\frac{z}{2}\tilde{X}_{2})\otimes \tilde{X}%
_{1}+\tilde{X}_{1}\otimes \exp (\frac{z}{2}\tilde{X}_{2}),  \notag \\
\Delta (\tilde{X}_{2}) &=&\exp (-\frac{z}{2}\tilde{X}_{2})\otimes \tilde{X}%
_{3}+\tilde{X}_{3}\otimes \exp (\frac{z}{2}\tilde{X}_{2}),  \notag
\end{eqnarray}%
satisfying the main homomorphism property for the whole deformed universal
enveloping algebra $U_{z}(so(2,1)\mathcal{)}.$

\QTP{Body Math}
Consider now some realization of the deformed generators $\tilde{X}_{i}\in
U_{z}(\mathcal{G}),$ $i=\overline{1,n},$ that is a homomorphism mapping $%
D_{z}:U_{z}(\mathcal{G})\rightarrow \mathcal{P(}M\mathcal{)},$ such that 
\begin{equation}
D_{z}(\tilde{X}_{i})=\tilde{e}_{i},  \label{3.9}
\end{equation}
$i=\overline{1,n},$ are some elements of a Poisson manifold $\mathcal{P(}M%
\mathcal{)}$ realized as a space of functions on a finite-dimensional
manifold $M,$ satisfying the deformed commutator relationships%
\begin{equation}
\{\tilde{e}_{i},\tilde{e}_{j}\}_{\mathcal{P(}M\mathcal{)}}=\vartheta _{z,ij}(%
\tilde{e}),  \label{3.10}
\end{equation}%
where, by definition, expressions $[\tilde{X}_{i},\tilde{X}_{j}]=\vartheta
_{z,ij}(\tilde{X}),$ $i,j=\overline{1,n},$ generate a Poisson co-algebra
structure on the function space $\mathcal{P}_{z}:=\mathcal{P}_{z}(\mathcal{G}%
)$ linked with a given Lie algebra $\mathcal{G}.$ Making \ \bigskip use of
the homomorphism property (\ref{3.3}) for the coproduct mapping $\Delta :%
\mathcal{P}_{z}(\mathcal{G})\rightarrow \mathcal{P}_{z}(\mathcal{G})\otimes 
\mathcal{P}_{z}(\mathcal{G}),$ one finds that for all $i,j=\overline{1,n}$ 
\begin{equation}
\{\Delta (\tilde{x}_{i}),\Delta (\tilde{x}_{j})\}_{\mathcal{P}_{z}(\mathcal{G%
})\otimes \mathcal{P}_{z}(\mathcal{G})}=\Delta (\{\tilde{x}_{i},\tilde{x}%
_{j}\}_{\mathcal{P}_{z}(\mathcal{G})}=\vartheta _{z,ij}(\Delta (\tilde{x}))
\label{3.11}
\end{equation}%
and for the corresponding coproduct $\Delta :\mathcal{P}(M)\rightarrow 
\mathcal{P}(M)\otimes \mathcal{P}(M)$ one gets similarly%
\begin{equation}
\{\Delta (\tilde{e}_{i}),\Delta (\tilde{e}_{j})\}_{\mathcal{P}(M)\otimes 
\mathcal{P}(M)}=\Delta (\{\tilde{e}_{i},\tilde{e}_{j}\}_{\mathcal{P}%
(M)}=\vartheta _{z,ij}(\Delta (\tilde{e})),  \label{3.12}
\end{equation}%
where \bigskip $\{.,.\}_{\mathcal{P}(M)}$ is some, eventually, canonical
Poisson structure on a finite-dimensional manifold $M.$ \ 

\QTP{Body Math}
Let $q\in M$ be a point of $M$ and consider its coordinates as elements of $%
\mathcal{P}(M).$ Then one can define the following elements 
\begin{equation}
q_{j}:=(I\otimes )^{j-1}q(\otimes I)^{N-j}\in \overset{(N)}{\otimes }%
\mathcal{P}(M),  \label{3.13}
\end{equation}%
where $j=\overline{1,N}$ \ by means of which one can construct the
corresponding $N$-tuple realization of the Poisson co-algebra structure (\ref%
{3.12}) as follows:%
\begin{equation}
\{\tilde{e}_{i}^{(N)},\tilde{e}_{j}^{(N)}\}_{\overset{(N)}{\otimes }\mathcal{%
P}(M)}=\vartheta _{z,ij}(\tilde{e}^{(N)}),  \label{3.14}
\end{equation}%
with $i,j=\overline{1,n}$ and 
\begin{equation}
\overset{(N)}{\otimes }D_{z}(\Delta ^{(N-1)}(\tilde{e}_{i}):=\tilde{e}%
_{i}^{(N)}(q_{1},q_{2},...,q_{N}).  \label{3.15}
\end{equation}%
For instance, for the $U_{z}(so(2,1)\mathcal{)}$ case (\ref{3.6}), one can
take \cite{BR} the realization Poisson manifold $\mathcal{P}(M)=\mathcal{P}(%
\mathbb{R}^{2})$ with the standard canonical Heisenberg-Weil Poissonian
structure on it:%
\begin{equation}
\{q,q\}_{\mathcal{P}(\mathbb{R}^{2})}=0=\{p,p\}_{\mathcal{P}(\mathbb{R}%
^{2})},\text{ \ \ \ \ \ \ \ }\{p,q\}_{\mathcal{P}(\mathbb{R}^{2})}=1,
\label{3.16}
\end{equation}%
where $(q,p)\in \mathbb{R}^{2}.$ Then expressions (\ref{3.15}) for $N=2$
give rise to the following relationships%
\begin{equation}
:%
\begin{array}{ccc}
\tilde{e}_{1}^{(2)}(q_{1},q_{2},p_{1},p_{2}) & := & (D_{z}\otimes
D_{z})\Delta (\tilde{X}_{1})= \\ 
2\frac{\sinh (\frac{z}{2}p_{1})}{z}\cos q_{1}\exp (\frac{z}{2}p_{1}) & + & 
2\exp (-\frac{z}{2}p_{1}))\frac{\sinh (\frac{z}{2}p_{2})}{z}\cos q_{2}, \\ 
\tilde{e}_{2}^{(2)}(q_{1},q_{2},p_{1},p_{2}) & := & (D_{z}\otimes
D_{z})\Delta (\tilde{X}_{2})=p_{1}+p_{2}, \\ 
\tilde{e}_{3}^{(2)}(q_{1},q_{2},p_{1},p_{2}) & := & (D_{z}\otimes
D_{z})\Delta (\tilde{X}_{3})= \\ 
2\frac{\sinh (\frac{z}{2}p_{1})}{z}\sin q_{1}\exp (\frac{z}{2}p_{2}) & + & 
2\exp (-\frac{z}{2}p_{1}))\frac{\sinh (\frac{z}{2}p_{2})}{z}\sin q_{2},%
\end{array}
\label{3.17}
\end{equation}%
where elements $(q_{1},q_{2},p_{1},p_{2})\in \mathcal{P}(\mathbb{R}%
^{2})\otimes \mathcal{P}(\mathbb{R}^{2})$ satisfy the induced by (\ref{3.16}%
) Heisenberg-Weil commutator relations:

\begin{equation}
\{q_{i},q_{j}\}_{\mathcal{P}(\mathbb{R}^{2})\otimes \mathcal{P}(\mathbb{R}%
^{2})}=0=\{p_{i},p_{j}\}_{\mathcal{P}(\mathbb{R}^{2})\otimes \mathcal{P}(%
\mathbb{R}^{2})},\text{ \ }\{p_{i},q_{j}\}_{\mathcal{P}(\mathbb{R}%
^{2})\otimes \mathcal{P}(\mathbb{R}^{2})}=\delta _{ij}  \label{3.18}
\end{equation}%
for any $i,j=\overline{1,2}.$

\section{$\protect\bigskip $Casimir elements and the Heisenberg-Weil algebra
related algebraic structures}

\setcounter{equation}{0}Consider any Casimir element $\tilde{C}\in U_{z}(%
\mathcal{G})$ related with an $\mathbb{R}\ni z-$deformed Lie algebra $%
\mathcal{G}$ structure in the form%
\begin{equation}
\lbrack \tilde{X}_{i},\tilde{X}_{j}]=\vartheta _{z,ij}(\tilde{X}),
\label{4.1}
\end{equation}%
where $i,j=\overline{1,n},$ $n=\dim \mathcal{\mathcal{G}}$, and, by
definition, $[\tilde{C},\tilde{X}_{i}]=0.$ The following general lemma holds.

\begin{lemma}
Let $(U_{z}(\mathcal{G});\Delta )$ be a co-algebra with generators
satisfying (\ref{4.1}) and $\tilde{C}\in U_{z}(\mathcal{G})$ be its Casimir
element; \ then%
\begin{equation}
\lbrack \Delta ^{(m)}(\tilde{C}),\Delta ^{(N)}(\tilde{X}_{i})]_{\overset{%
(N+1)}{\otimes }U_{z}(\mathcal{G})}=0  \label{4.2}
\end{equation}%
for any $i=\overline{1,n}$ and $m=\overline{1,N}.$
\end{lemma}

As a simple corollary of this Lemma one finds from (\ref{4.2}) that $\ $%
\begin{equation*}
\lbrack \Delta ^{(m)}(\tilde{C}),\Delta ^{(N)}(\tilde{C})]_{\overset{(N+1)}{%
\otimes }U_{z}(\mathcal{G})}=0
\end{equation*}
for any $k,m\in \mathbb{Z}_{+}.$

Consider now some realization (\ref{3.9}) of our deformed \ Poisson
co-algebra structure (\ref{4.1}) and check that the expression 
\begin{equation}
\lbrack \Delta ^{(m)}(C(\tilde{e}),\Delta ^{(N)}(\mathcal{H(}\tilde{e}%
\mathcal{)})]_{\overset{(N+1)}{\otimes }\mathcal{P}(M)}=0  \label{4.3}
\end{equation}%
too for any $m=\overline{1,N},$ $N\in \mathbb{Z}_{+},$ if $C(\tilde{e})\in I(%
\mathcal{P}(M)),$ that is $\{C(\tilde{e}),q\}_{\mathcal{P}(M)}=0$ for any $%
q\in M.$ Since 
\begin{equation}
\mathcal{H}^{(N)}(q):=\Delta ^{^{(N-1)}}(\mathcal{H(}\tilde{e}\mathcal{)})
\label{4.5}
\end{equation}%
are in general, smooth functions on $\overset{(N+1)}{\otimes }M,$ which can
be used as Hamilton ones subject to the Poisson structure on $\overset{(N+1)}%
{\otimes }\mathcal{P}(M),$ the expressions (\ref{4.5}) mean nothing else
that functions 
\begin{equation}
\gamma ^{(m)}(q):=\Delta ^{^{(N)}}(C\mathcal{(}\tilde{e}\mathcal{)})
\label{4.6}
\end{equation}%
are their invariants, that is 
\begin{equation}
\{\gamma ^{(m)}(q),\mathcal{H}^{^{(N)}}(q)\}_{\overset{(N+1)}{\otimes }%
\mathcal{P}(M)}=0  \label{4.7}
\end{equation}%
for any $m=\overline{1,N}.$ Thereby, the functions (\ref{4.5}) and (\ref{4.6}%
) generate under some additional but natural conditions a hierarchy of a
priori Liouville-Arnold integrable Hamiltonian flows on the Poisson manifold 
$\overset{(N+1)}{\otimes }\mathcal{P}(M).$

Consider now a case when a Poisson manifold $\mathcal{P}(M)$ and its
co-algebra deformation $\mathcal{P}_{z}(\mathcal{G}).$ Thus for any
coordinate points $x_{i}\in \mathcal{P}(\mathcal{G}),$ $i=\overline{1,n},$
the following relationships 
\begin{equation}
\{x_{i},x_{j}\}=\tsum\limits_{k=1}^{n}c_{ij}^{k}x_{k}:=\vartheta _{ij}(x)
\label{4.8}
\end{equation}%
define a Poisson structure on $\mathcal{P}(\mathcal{G}),$ related with the
corresponding Lie algebra structure of $\mathcal{G},$ and there exists a
representation (\ref{3.9}), such that elements $\tilde{e}_{i}:=D_{z}(\tilde{X%
}_{i})=\tilde{e}_{i}(x)$ satisfy the relationships $\{\tilde{e}_{i},\tilde{e}%
_{j}\}_{\mathcal{P}_{z}(\mathcal{G})}=\vartheta _{z,ij}(\tilde{e})$ for any $%
i=\overline{1,n},$ with the limiting conditions 
\begin{equation}
\underset{z\rightarrow 0}{\lim }\vartheta _{z,ij}(\tilde{e}%
)=\tsum\limits_{k=1}^{n}c_{ij}^{k}x_{k},\text{ \ \ }\underset{z\rightarrow 0}%
{\lim }\tilde{e}_{i}(x)=x_{i}  \label{4.9}
\end{equation}%
for any $i,j=\overline{1,n}$ being held. For instance, take the Poisson
co-algebra $\mathcal{P}_{z}(so(2,1))$ for which there exists a realization (%
\ref{3.9}) in the following form:%
\begin{eqnarray}
\tilde{e}_{1} &:&=D_{z}(\tilde{X}_{1})=\frac{\sinh (\frac{z}{2}x_{2})}{zx_{2}%
}x_{1},\text{ }\tilde{e}_{2}:=D_{z}(\tilde{X}_{2})=x_{2},  \label{4.10} \\
\tilde{e}_{3} &:&=D_{z}(\tilde{X}_{3})=\frac{\sinh (\frac{z}{2}x_{2})}{zx_{2}%
}x_{3},  \notag
\end{eqnarray}%
where $x_{i}\in \mathcal{P}(so(2,1)),$ $i=\overline{1,3},$ satisfy the $%
so(2,1)-$commutator relationships%
\begin{eqnarray}
\{x_{2},x_{1}\}_{\mathcal{P}(so(2,1))} &=&x_{3},\text{ \ }\{x_{2},x_{3}\}_{%
\mathcal{P}(so(2,1))}=-x_{1},  \label{4.11} \\
\{x_{3},x_{1}\}_{\mathcal{P}(so(2,1))} &=&x_{2,}  \notag
\end{eqnarray}%
with the coproduct operator $\Delta :\bigskip \bigskip \mathcal{U}%
_{z}(so(2,1))\rightarrow \mathcal{U}_{z}(so(2,1))\otimes \mathcal{U}%
_{z}(so(2,1))$ being given by (\ref{3.8}). It is easy to check that
conditions (\ref{4.8}) and (\ref{4.9}) hold.

The next example is related with the co-algebra $\mathcal{U}_{z}(\pi (1,1))$
of the Poincare algebra $\pi (1,1)$ for which the following non-deformed
relationships 
\begin{equation}
\lbrack X_{1},X_{2}]=X_{3},\text{ }[X_{1},X_{3}]=X_{2},\text{ }%
[X_{3},X_{2}]=0  \label{4.12}
\end{equation}%
hold. The corresponding coproduct $\Delta :\mathcal{U}_{z}(\pi
(1,1))\rightarrow \mathcal{U}_{z}(\pi (1,1))\otimes \mathcal{U}_{z}(\pi
(1,1))$ is given by the Woronowicz \ \cite{Wo} expressions%
\begin{eqnarray}
\Delta (\tilde{X}_{1}) &=&I\otimes \tilde{X}_{1}+\tilde{X}_{1}\otimes I,
\label{4.13} \\
\Delta (\tilde{X}_{2}) &=&\exp (-\frac{z}{2}\tilde{X}_{1})\otimes \tilde{X}%
_{1}+\tilde{X}_{1}\otimes \exp (\frac{z}{2}\tilde{X}_{1}),  \notag \\
\Delta (\tilde{X}_{3}) &=&\exp (-\frac{z}{2}\tilde{X}_{1})\otimes \tilde{X}%
_{3}+\tilde{X}_{3}\otimes \exp (\frac{z}{2}\tilde{X}_{1}),  \notag
\end{eqnarray}%
where $z\in \mathbb{R}$ is a prarameter. Under the deformed expressions (\ref%
{4.13}) the elements $\tilde{X}_{j}\in \mathcal{U}_{z}(\pi (1,1)),$ $j=%
\overline{1,3},$satisfy still undeformed commutator relationships, that is $%
\vartheta _{z,ij}(\tilde{X})=\left. \vartheta _{ij}(X)\right\vert
_{X\Rightarrow \tilde{X}}$ for any $z\in \mathbb{R},$ $i,j=\overline{1,3},$
being given by (\ref{4.12}). As a result, we can state that $\ \tilde{e}%
_{i}:=D_{z}(\tilde{X}_{i})=\tilde{e}_{i}(x)=x_{i},$ where for $x_{i}\in 
\mathcal{P}(\pi (1,1)),$ $i=\overline{1,3},$ the following Poisson structure 
\begin{eqnarray}
\{x_{1},x_{2}\}_{\mathcal{P}(\pi (1,1))} &=&x_{3},\text{ \ }\{x_{1},x_{3}\}_{%
\mathcal{P}(\pi (1,1))}=x_{2},  \label{4.14} \\
\{x_{3},x_{2}\}_{\mathcal{P}(\pi (1,1))} &=&0  \notag
\end{eqnarray}%
holds. Moreover, since $C=x_{2}^{2}-x_{3}^{2}\in I(\mathcal{P}(\pi (1,1))),$
that is $\{C,x_{i}\}_{\mathcal{P}(\pi (1,1))}=0$ for any $i=\overline{1,3},$
on can construct, making use of (\ref{4.5}) and (\ref{4.6}), integrable
Hamiltonian systems on $\overset{(N)}{\otimes }\mathcal{P}(\pi (1,1)).$The
same one can do subject to the discussed above Poisson co-algebra $\mathcal{P%
}_{z}(so(2,1))$ realized by means of the Poisson manifold $\mathcal{P}%
(so(2,1)),$ taking into account that the following element $%
C=x_{2}^{2}-x_{1}^{2}-x_{3}^{2}\in I(\mathcal{P}(so(2,1)))$ is a Casimir one.

Now we will consider a special extended Heisenberg-Weil co-algebra $\mathcal{%
U}_{z}(h_{4}),$ called still the oscillator co-algebra. The undeformed Lie
algebra $h_{4}$ commutator relationships take the form:

\begin{eqnarray}
\text{ }[n,a_{+}] &=&a_{+},\text{ \ }[n,a_{-}]=-a_{-},  \label{4.15} \\
\lbrack a_{-},a_{+}] &=&m,\text{ \ \ }[m,\cdot ]=0,  \notag
\end{eqnarray}%
where $\{n,a_{\pm },m\}\bigskip \subset h_{4}$ compile a basis of $%
h_{4},\dim h_{4}=4.$ The Poisson co-algebra $\mathcal{P}(h_{4})$ is
naturally endowed with the Poisson structure like (\ref{4.15}) and admits
its realization (\ref{3.9}) on the Poisson manifold $\mathcal{P}(\mathbb{R}%
^{2}).$ Namely, on $\bigskip \mathcal{P}(\mathbb{R}^{2})$ one has 
\begin{eqnarray}
e_{\pm } &=&D(a_{\pm })=\sqrt{p}\exp (\mp q),\text{ \ }  \label{4.16} \\
e_{1} &=&D(m)=1,\text{ }e_{0}=D(n)=p,  \notag
\end{eqnarray}%
where $(q,p)\in \mathbb{R}^{2}$ and the Poisson structure on $\mathcal{P}(%
\mathbb{R}^{2})$ is canonical, that is the same as (\ref{3.16}).

Closely related with the relationships (\ref{4.15}) there is a generalized $%
\mathcal{U}_{z}(su(2))$ co-algebra, for which 
\begin{eqnarray}
\lbrack x_{3},x_{\pm }] &=&\pm x_{\pm },\text{ \ \ \ \ \ \ }[y_{\pm },\cdot
]=0,  \label{4.17} \\
\lbrack x_{+},x_{-}] &=&y_{+}\sin (2zx_{3})+y_{-}\cos (2zx_{3}))\frac{1}{%
\sin z},\text{ \ \ }  \notag
\end{eqnarray}%
where $z\in \mathbb{C}$ is an arbitrary parameter. The co-algebra structure
is given now as follows:%
\begin{eqnarray}
\Delta (x_{\pm }) &=&c_{1(2)}^{\pm }e^{izx_{3}}\otimes x_{\pm }+x_{\pm
}\otimes c_{2(1)}^{\pm }e^{-izx_{3}},  \label{4.18} \\
\Delta (x_{3}) &=&I\otimes x_{3}+x_{3}\otimes I,\text{ }\Delta (c_{i}^{\pm
})=c_{i}^{\pm }\otimes c_{i}^{\pm },  \notag \\
\nu (x_{\mp }) &=&-(c_{1(2)}^{\pm })^{-1}e^{-izx_{3}}x_{\mp
}e^{izx_{3}}(c_{2(1)}^{\pm })^{-1},  \notag \\
\nu (c_{i}^{\pm }) &=&(c_{i}^{\pm })^{-1},\text{ \ }\nu (e^{\pm
izx_{3}})=e^{\mp izx_{3}}  \notag
\end{eqnarray}%
with $c_{i}^{\pm }\in \mathcal{U}_{z}(su(2)),$ $i=\overline{1,2},$ being
fixed elements. One can check that the corresponding to (\ref{4.17}) Poisson
structure on $\mathcal{P}_{z}(su(2))$ can be realized by means of the
canonical Poisson structure on the phase space $\mathcal{P}(\mathbb{R}^{2})$
as follows:%
\begin{eqnarray}
\lbrack q,p] &=&i,\text{ \ \ \ \ \ \ \ \ }D_{z}(x_{3})=q,\ \ \ \text{\ \ \ \
\ }\ D_{z}(x_{\mp })=e^{\pm ip}g_{z}(q),  \label{4.19} \\
g_{z}(q) &=&(k+\sin [z(s-q)])(y_{+}\sin [(q+s+1)]+y_{-}\cos [z(q+s+1)])^{1/2}%
\frac{1}{\sin z},  \notag
\end{eqnarray}%
where $k,s\in \mathbb{C}$ are constant parameters. Thereby making use of (%
\ref{4.6}) and (\ref{4.7}), one can construct a new class of Liouville
integrable Hamiltonian flows.

\section{The Heisenberg-Weil co-algebra structure and related integrable
flows}

\setcounter{equation}{0}Consider the Heisenberg-Weil algebra commutator
relationships (\ref{4.15}) and related with them the fo9llowing homogenous
quadratic forms%
\begin{equation}
\left. 
\begin{array}{c}
x_{1}x_{2}-x_{2}x_{1}-\alpha x_{3}^{2}=0, \\ 
x_{1}x_{3}-x_{3}x_{1}=0,\text{ \ \ }x_{2}x_{3}-x_{3}x_{2}=0%
\end{array}%
\right\} R(x),  \label{5.1}
\end{equation}%
where $\alpha \in \mathbb{C},$ \ \ $x_{i}\in A,$ $i=\overline{1,3},$ are
some elements of a free associative algebra A. The quadratic algebra $A/R(x)$
can be deformed via 
\begin{equation}
\left. 
\begin{array}{c}
x_{1}x_{2}-z_{1}x_{2}x_{1}-\alpha x_{3}^{2}=0, \\ 
x_{1}x_{3}-z_{2}x_{3}x_{1}=0,\text{ \ \ }x_{2}x_{3}-z_{2}^{-1}x_{3}x_{2}=0,%
\end{array}%
\right\} R_{z}(x),  \label{5.2}
\end{equation}%
where $z_{1},z_{2}\in \mathbb{C}\backslash \{0\}$ are some parameters.

Let $V$ be the vector space of columns $X:=(x_{1},x_{2},x_{3})^{\intercal }$
and define the following action%
\begin{equation}
h_{T}:V\rightarrow (V\otimes V^{\ast })\otimes V,\text{ }  \label{5.3}
\end{equation}%
where, by definition, for any $X\in V$%
\begin{equation}
h_{T}(X)=T\otimes X.  \label{5.4}
\end{equation}%
It is easy to check that conditions (\ref{5.2}) will be satisfied if the
following relations \cite{BI} 
\begin{eqnarray}
T_{11}T_{33} &=&T_{33}T_{11},\text{ \ }T_{12}T_{33}=z_{2}^{-2}T_{33}T_{12},%
\text{ }T_{21}T_{33}=z_{1}^{2}T_{33}T_{21},  \label{5.5} \\
T_{22}T_{33} &=&T_{33}T_{22},\text{ \ }T_{31}T_{33}=z_{2}T_{33}T_{31},\text{
\ \ }T_{32}T_{33}=z_{1}^{-1}T_{33}T_{32},  \notag \\
T_{11}T_{12} &=&z_{1}T_{12}T_{11},\text{ }T_{21}T_{22}=z_{1}T_{22}T_{21},%
\text{ }z_{2}T_{11}T_{32}-z_{2}T_{32}T_{11}=  \notag \\
&=&z_{1}z_{2}T_{12}T_{31}-T_{31}T_{12},\text{ }%
T_{21}T_{32}-z_{1}z_{2}T_{32}T_{21}\text{ =}  \notag \\
&=&z_{1}T_{22}T_{31}-z_{2}T_{31}T_{22},\text{ \ }T_{11}T_{22}-T_{22}T_{11}= 
\notag \\
&=&z_{1}T_{12}T_{21}-z_{1}^{-1}T_{21}T_{12},\text{ \ (}%
T_{11}T_{22}-z_{1}T_{12}T_{21})=  \notag \\
&=&\alpha T_{33}^{2}-T_{31}T_{32}+z_{1}T_{32}T_{31}  \notag
\end{eqnarray}%
hold. Put now for further convenience $z_{1}=z_{2}^{2}:=z^{2}\in \mathbb{C}$
and compute the "quantum" determinant $D(T)$ of the matrix $%
T:(A/R_{z}(x))^{3}\rightarrow (A/R_{z}(x))^{3}:$%
\begin{equation}
D(T)=(T_{11}T_{22}-z^{-2}T_{21}T_{12})T_{33}.  \label{5.6}
\end{equation}%
Remark here that the determinant (\ref{5.6}) is not central, that is 
\begin{eqnarray}
D^{-1}T_{11} &=&T_{11}D^{-1},\text{ \ }D^{-1}T_{12}=z^{-6}T_{12}D^{-1},
\label{5.7} \\
D^{-1}T_{33} &=&T_{33}D^{-1},\text{ }\ z^{-6}D^{-1}T_{21}=T_{12}D^{-1}, 
\notag \\
D^{-1}T_{22} &=&T_{22}D^{-1},\text{ \ }z^{-3}D^{-1}T_{31}=T_{31}D^{-1},\text{
\ }  \notag \\
D^{-1}T_{32} &=&z^{-3}T_{32}D^{-1}.  \notag
\end{eqnarray}%
Taking into account properties (\ref{5.5}) - (\ref{5.7}), one can construct
the Heisenberg-Weil related co-algebra $\mathcal{U}_{z}(h)$ being a Hopf
algebra with the fo9llowing coproduct $\Delta ,$ counit $\varepsilon $ and
antipode $\nu :$%
\begin{eqnarray}
\Delta (T) &:&=T\otimes T,\text{ \ }\Delta (D^{-1}):=D^{-1}\otimes D^{-1},
\label{5.8} \\
\varepsilon (T) &:&=I,\text{ \ }\varepsilon (D^{-1}):=I,\text{ \ }\nu
(T):=T^{-1},\text{ }\nu (D):=D^{-1}.  \notag
\end{eqnarray}

Based now on relationships (\ref{5.5}), one can easily construct the Poisson
tensor 
\begin{equation}
\{\Delta (\tilde{T}),\Delta (\tilde{T})\}_{\mathcal{P}_{z}(h)\otimes 
\mathcal{P}_{z}(h)}=\Delta (\{\tilde{T},\tilde{T}\}_{\mathcal{P}%
_{z}(h)}):=\vartheta _{z}(\Delta (\tilde{T})),  \label{5.9}
\end{equation}%
subject to which all of functionals (\ref{4.6}) will be commuting to each
other, and moreover, will be Casimir ones. Choosing some appropriate
Hamiltonian functions $\mathcal{H}^{(N)}(\tilde{T}):=\Delta ^{(N-1)}(%
\mathcal{H}(\tilde{T}))$ for $N\in \mathbb{Z}_{+}$ one makes it possible to
present a priori nontrivial integrable Hamiltonian systems. On the other
handside, the co-algebra $\mathcal{U}_{z}(h)$ built \ by (\ref{5.7}) and (%
\ref{5.8}) possesses the following fundamental $\mathcal{R}$-matrix \cite%
{KBI} property:%
\begin{equation}
\mathcal{R}(z)(T\otimes I)(I\otimes T)=(I\otimes T)(T\otimes I)\mathcal{R}(z)
\label{5.10}
\end{equation}%
for some complex-valued matrix $\mathcal{R}(z)\in Aut(\mathbb{C}^{3}\otimes 
\mathbb{C}^{3}),$ $z\in \mathbb{C}.$ The latter, as is well known \cite{KBI}%
, gives rise to a regular procedure of constructing an infinite hierarchy of
Liouville-integrable operator (quantum) Hamiltonian systems on related
quantum Poissonian phase spaces. \ On their special cases interesting for
applications we plan to go on in another place.

\end{document}